\documentclass[showpacs,preprint,superscriptaddress]{revtex4}
\usepackage{graphicx}

\begin{document}

\title{Magnetic Diode Effect in Double Barrier Tunnel Junctions}

\author{M. Chshiev}
\affiliation{Institut de Physique et Chimie des Matériaux de
Strasbourg (UMR 7504 CNRS-ULP), 23 rue du Loess, 67037 Strasbourg,
France}
\author{D. Stoeffler}
\affiliation{Institut de Physique et Chimie des Matériaux de
Strasbourg (UMR 7504 CNRS-ULP), 23 rue du Loess, 67037 Strasbourg,
France}
\author{A. Vedyayev}
\affiliation{Faculty of Physics, Moscow Lomonosov State
University, Moscow, 119899 Russia}
\author{K. Ounadjela}
\affiliation{Institut de Physique et Chimie des Matériaux de
Strasbourg (UMR 7504 CNRS-ULP), 23 rue du Loess, 67037 Strasbourg,
France}

\begin{abstract}
A quantum statistical theory of spin-dependent tunneling through
asymmetric magnetic double barrier junctions is presented which
describes $both$ ballistic and diffuse tunneling by a single
analytical expression. It is evidenced that the key parameter for
the transition between these two tunneling regimes is the electron
scattering. For these junctions a strong asymmetric behaviour in
the \mbox{$I$-$V$} characteristics and the tunnel
magnetoresistance (TMR) is predicted which can be controlled by an
applied magnetic field. This phenomenon relates to the quantum
well states in the middle metallic layer. The corresponding
resonances in the current and the TMR are drastically phase
shifted under positive and negative voltage.
\end{abstract}

\maketitle

After the discovery of a large room temperature magnetoresistance
effect in magnetic tunnel junctions (MTJ)~\cite{Moodera1} many
potential applications have emerged based on the spin polarized
transport through a thin insulating barrier sandwiched between two
ferromagnetic metals\cite{Moodera1,Moodera2,Miyazaki,Tezuka}. The
current state of the art production of tunnel junction elements
with a resistance variation of up to 40\% when magnetically
switched~\cite{Parkin,Freitas} makes them very promising in
particular for the application as tunneling magnetic random access
memories (MRAM). Current MRAM designs~\cite{Parkin,Tehrani} which
incorporate arrays of magnetic tunnel junctions add an additional
semiconductor switch in series (either a CMOS transistor or a
$p$-$n$ junction) with the MTJ memory cell to suppress (or block)
parasitic signal paths within the array of lines. However, such a
concept is hampered by the difficulty of mixing semiconductor and
metal technology. In this letter, we propose a novel theoretical
concept which eliminates the introduction of additional
semiconductor components by using an asymmetric double barrier
structure \mbox{M$_1$/O$_2~a$/M$_3~b$/O$_4~c$/M$_5$} (M$_i$ and
O$_i$ are the magnetic metal and oxide layer respectively with
corresponding thickness $a$, $b$ and $c$) that in itself acts as
the blocking device. It is shown that the \mbox{$I$-$V$}
characteristics and the TMR of such asymmetric double barrier
structures have a strong "diode"-like behaviour under positive and
negative applied voltage. Moreover, the asymmetric properties can
be varied by a magnetic field leading to the concept of a
"magnetically controlled" diode.

The potential profile of the system under applied voltage
$\mathrm{V_{ext}}$ is shown in Fig.~1(a), where $U_i$ and
${\mathrm{V}}_i$ ($i$=2,4) are respectively the potential of the
barrier region and the linear voltage drop therein, and
$V_i^\sigma$ ($i$=1,3,5) is the spin-dependent potential of the
$i$-th metal. The outer metallic layers are assumed to be
semi-infinite. The symmetric structures with $a=c$ and $U_2=U_4$
were considered in references~\cite{Xiangdong} and~\cite{Barnas}
but the resonance tunneling was described without taking into
account the electron's scattering inside the middle metallic
layer. In this case it was shown that the conductivity and the TMR
exhibit resonance peaks as a function of the thickness of the
middle metallic layer due to quantum well states. Here we will
show that the asymmetry of the structure in combination with the
presence of quantum well states leads to a large asymmetry in the
current for forward (positive) and reverse (negative) applied
voltage.
\begin{figure}
\begin{center}
\includegraphics[width=0.95\textwidth,angle=0]{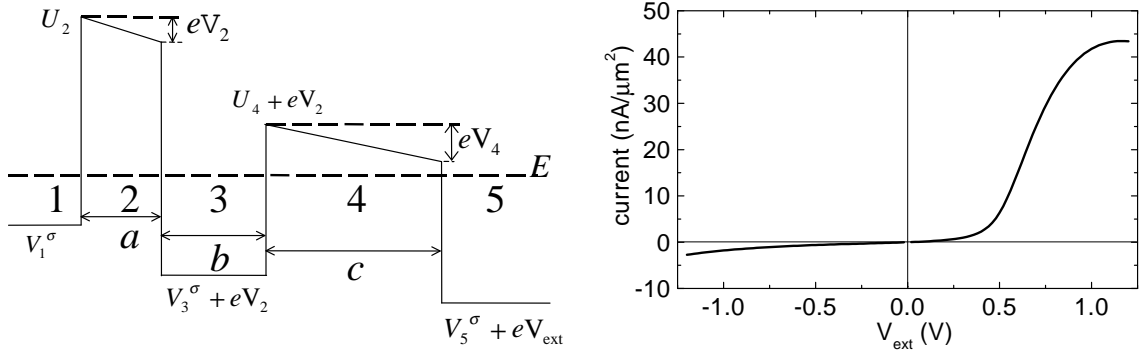}
\caption{(a) Potential energy diagram and (b) the calculated
current-voltage curve for the asymmetric structure \mbox{Cu/O$_2$
7~\AA/Cu 5.5~\AA/O$_4$ 21~\AA/Cu} with
$U_2-E_F=U_4-E_F=3~e\mathrm{V}$.}
\end{center}
\end{figure}
The evaluation of the current through the double barrier junction
is based on the determination of its transmission probability
$D$~\cite{Duke}. Here, $D$ is derived using the Green functions
technique in the mixed real space-momentum
representation~\cite{Vedyayev2,Fisher,Camblong} and it can be
written in the form:
\begin{equation}
\label{probformula}
\begin{array}{cr} \displaystyle D=
{\Bigg({\hbar^2\over{2m}}\Bigg)^2}
&\displaystyle\Bigg\{\left[{\partial{G_\kappa}\over{\partial{z}}}-
{\partial{G_\kappa^*)}\over{\partial{z}}}\right]
\left[{\partial{G_\kappa}\over{\partial{z^\prime}}}-
{\partial{G_\kappa^*}\over{\partial{z^\prime}}}\right]-
\displaystyle
\left[{\partial^2{G_\kappa}\over{\partial{z}\partial{z^\prime}}}-
{\partial^2{G_\kappa^*}\over{\partial{z}\partial{z^\prime}}}\right]
\bigg[{G_\kappa}-{G_\kappa^*}\bigg]\Bigg\}
\end{array}
\end{equation}
where $G_\kappa\equiv{G_\kappa}(z,z^\prime)$ and
$G_\kappa^*\equiv{G_\kappa^*}(z,z^\prime)$ are retarded and
advanced Green functions of the system, $z$ is the coordinate
perpendicular to the plane of the structure and $\kappa$ indicates
the in-plane momentum of the electron. This expression is similar
to the Kubo formula for the non-local conductivity~\cite{Fisher}
by replacing Fermi energy $E_F$ with an arbitrary energy $E$ over
which an integration has to be done in the limits from $E_F$ upto
$E_F+e{\mathrm{V}}_{ext}$. It is convenient to use
eq.~(\ref{probformula}) since for the calculation of the Green
function of a system including scattering it is possible to use
well defined quantum statistical methods. In the presence of
scattering in the metallic layers, the problem can not be solved
exactly and the tunneling through the whole system is described by
consecutive tunneling through each barrier. In this case the
current densities in the first and second barrier are calculated
separately and the condition of constant current throughout the
structure has to be fulfilled by introducing either an effective
electrical field~\cite{Camblong,Levy} inside each barrier or by
calculating so called vertex correction~\cite{Camblong,Kane}. Here
the first approach is used. The Green function is the solution to
the Schr\"odinger equation in each layer, where in the barrier
region the WKB approximation~\cite{Vedyayev4} is used: $$
\Bigg(E+{{\hbar^2}\over{2m_i}}\left({\partial^2\over{\partial{z}^2}}-{\kappa^2}\right)-
V_i^\sigma+i{2k^\sigma_{Ei}\over{l_i^\sigma}}-e{\mathrm{V}}_i(z)\Bigg)G_\kappa(z,z^\prime)=\delta(z-z^\prime),
\quad \left(i=1,3,5\right) $$ and $$
\Bigg(E+{{\hbar^2}\over{2m_i}}\left({\partial^2\over{\partial{z}^2}}-{\kappa^2}\right)-
\left(U_i+e
{\mathrm{V}}_i(z)\right)\Bigg)G_\kappa(z,z^\prime)=\delta(z-z^\prime),
\quad \left(i=2,4\right) $$ where
$k^\sigma_{Ei}=\sqrt{({2m_i/\hbar^2})(E-V_i^\sigma)}$,
$l_i^\sigma$ and $m_i$ are the mean free path of the electron with
energy $E$ and spin $\sigma$ and its effective mass in the $i$-th
layer, respectively, and voltage $${\mathrm{V}}_i(z)=\left\{
\begin{array}{ll}
0&z<0 \\ (z\mathrm{V}_2)/a&0<z<a \\ \mathrm{V}_2&a<z<a+b \\
\mathrm{V}_2+(z-a-b)\mathrm{V}_4/c&a+b<z<a+b+c \\
\mathrm{V}_{ext}&z>a+b+c
\end{array}\right.$$
The solutions in each layer are matched by the boundary condition
at the interfaces $z=0$, $a$, $a+b$, $a+b+c$~(see Fig.~1(a)). From
the Green function the non-local probability $D$ is calculated
using the formula~(\ref{probformula}). With the assumption that
the voltage drop occurs only across the barrier regions
($\mathrm{V}_2$ and $\mathrm{V}_4$ in Fig.~1(a)) and that the
conduction band edge is flat in the metallic regions, it is
sufficient to consider only the non-local probability for those
$z$ and $z^\prime$ lying in the barrier regions 2,
4~\cite{Vedyayev2} (see Fig.~1(a)). This yields the four
analytical expressions: $D_{22}$, $D_{24}$, $D_{42}$ and $D_{44}$.
It is found that these quantities are independent of $z$ and
$z^\prime$ and hence the condition that the divergence of the
current density is zero is automatically satisfied inside the
different regions. With these two point tunneling probabilities
the current density throughout the first ($j_2$) and the second
($j_4$) barrier can be calculated separately and are written in
the following form ref.~\cite{Duke}
\begin{equation}
\label{eq.3}
\begin{array}{ll}
j_{2(4)}^\sigma&=\displaystyle {e\over{\pi h}}\int dE
\big[f(E)-f(E+e\mathrm{V}_2)\big]\int
D_{22(42)}^\sigma(E,\kappa)\kappa d\kappa  \\
\\
&\displaystyle +{e\over{\pi h}}\int dE
\big[f(E+e\mathrm{V}_2)-f(E+e\mathrm{V}_{ext})\big]\int
D_{24(44)}^\sigma(E,\kappa)\kappa d\kappa
\end{array}
\end{equation}
where $f(E)$ is the thermal occupation probability of a state with
energy $E$. This is the generalization of the linear response in
metals where the current density at a point $z$ is related to the
electric field at a point $z^\prime$ through the two-point
conductivity $\sigma(z,z^\prime)$. Finally, from the requirement
that the current density has to be constant throughout the whole
barrier structure, $j_2^\sigma=j_4^\sigma$, the effective electric
field inside each barrier and hence the voltage drops
$\mathrm{V_2}$ and $\mathrm{V_4}$ are determined in a self
consistent way for a given applied voltage $\mathrm{V_{ext}}$. The
additional condition of the current conservation here results from
introducing the effective electric field instead of calculating
the vertex corrections~\cite{Levy}. Alternatively, if one
calculates the current taking into account the vertex correction,
the matching of the boundary conditions would be enough to provide
the current conservation. The resulting dependence of the current
density on $\mathrm{V_{ext}}$ is given in Fig.~1(b) for the case
of an asymmetric double barrier structure $a \ne c$ or $U_2 \ne
U_4$, revealing a strong asymmetry in the \mbox{$I$-$V$}
characteristic reminiscent of a diode.
\begin{figure}
\begin{center}
\includegraphics[width=0.95\textwidth,angle=0]{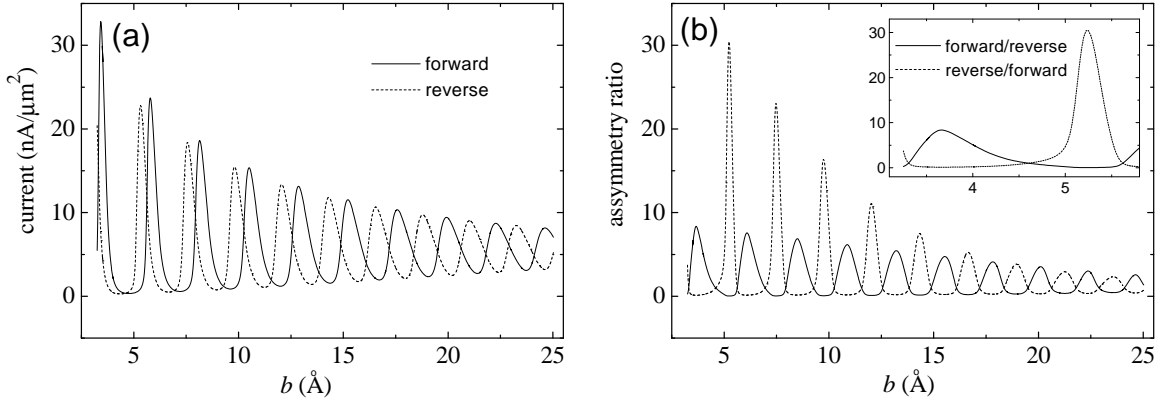}
\caption{Dependence of (a) the forward and reverse current and (b)
the corresponding asymmetry ratio on the layer thickness $b$ of
the middle metallic layer for the non-magnetic asymmetric double
barrier structure (see text).}
\end{center}
\end{figure}
To understand the physical origin of this enhanced asymmetry, we
first consider the case where M$_1$, M$_3$ and M$_5$ are
non-magnetic metals, using for the Fermi wave vector the value of
Cu $k_{E_Fi}=1.36$~\AA$^{-1}$~\cite{Kittel}. Cu is chosen here
because of its simple electronic structure similar to free
electrons. Furthermore, we use the electron mean free path of
100~\AA \ for the scattering in the metallic layers and an
electron effective mass of $m_{2(4)}$=$0.4$~\cite{Bratkovsky}
inside the barriers. In Fig.~2(a) the current density is shown as
a function of the thickness $b$ of the middle layer M$_3$ for a
fixed applied voltage $\mathrm{V}_{ext}=0.7~\mathrm{V}$ for a
double barrier structure of \mbox{Cu/O$_2$ 21~\AA/Cu $b$~\AA/O$_4$
21~\AA/Cu} with different barrier heights ($U_2-E_F=1~e\mathrm{V}$
and $U_4-E_F=3~e\mathrm{V}$). Both forward (positive
$\mathrm{V}_{ext}$) and reverse (negative $\mathrm{V}_{ext}$)
currents exhibit resonance peaks (oscillations) which are
associated with the formation of quantum well states in the middle
metal layer
3~\cite{Xiangdong,Barnas,Vedyayev2,Vedyayev4,Zhang,Mathon,Vedyayev3,Moodera3}.
The period of these oscillations is the same for both curves and
proportional to $\pi/k_{E_F}$ (see, for example,
ref.~\cite{Vedyayev2}), but the positions (phases) of the resonant
peaks are shifted with respect to each other. The phase is defined
by the boundary conditions at the metal/oxide interfaces. In the
case of asymmetric double barrier structures, the matching of the
phases at the interfaces is sensitive to the direction of the
current due to different $D_{22}$ and $D_{44}$ in (\ref{eq.3})
which depend exponentially on the barrier parameters leading to
asymmetric voltage drops $\mathrm{V_2}$ and $\mathrm{V_4}$. The
difference in these voltage drops will bring the quantum well
states in M$_3$ to line up differently with respect to the energy
$E$ of the electron under positive and negative applied voltage.
The resulting phase shift of the current density leads, thus, to a
current asymmetry ratio (forward current divided by reverse
current and vice versa) which oscillates with the same period as
the current density and which is considerably enhanced at its
maxima.
\begin{figure}
\begin{center}
\includegraphics[width=0.95\textwidth,angle=0]{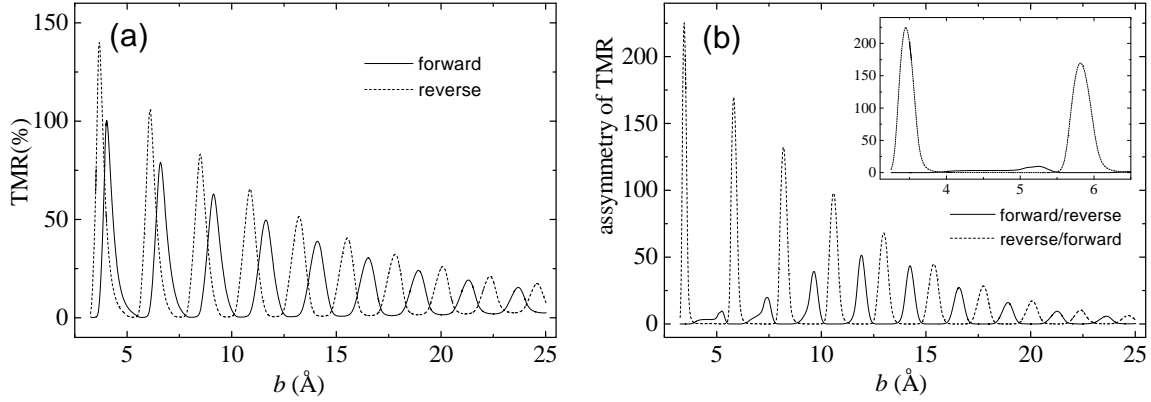}
\caption{Dependence of (a) the TMR ratio and (b) its corresponding
asymmetry ratio on $b$ calculated from the forward and reverse
current through an asymmetric double barrier structure with
magnetic outer layers (see text).}
\end{center}
\end{figure}
This oscillation of the asymmetry ratio as a function of $b$ is
shown in Fig.~2(b). Hence, choosing the appropriate parameters of
the layer thicknesses and the barrier heights, the asymmetry can
be enhanced significantly (more than one order of magnitude),
leading to the \mbox{$I$-$V$} characteristics presented in
Fig.~1(b) reminiscent of a diode. Similar characteristics were
found for double barrier structures with the same barrier heights
but different barrier thickness. In contrast, for a symmetric
double barrier structure ($U_2=U_4$ and $a=c$), the phase shift in
the forward and reverse current densities is zero, resulting in a
symmetric \mbox{$I$-$V$} curve for positive and negative applied
voltage. Consequently the diode behavior is lost.

Replacing in the asymmetric double barrier structure the outer
layers M$_1$ and M$_5$ by ferromagnetic metals, it is found that
the diode efficiency can be controlled by an applied magnetic
field. Furthermore, it is found that the tunneling
magnetoresistance (TMR) ratio itself depends strongly on the
direction of the current yielding a high asymmetry ratio. The
current density of such a magnetic double barrier structure was
calculated for the case of identical magnetic layers using Fermi
wave vectors $k_{E_F}^\uparrow=1.09$~\AA$^{-1}$ and
$k_{E_F}^\downarrow=0.42$~\AA$^{-1}$ which correspond to the
spin-split free-electron-like $d$-electron bands of
Fe~\cite{Stearns}. The parameters for the other layers are the
same as for the non-magnetic case discussed above. The TMR ratio
is defined as $$TMR={\sum_\sigma
(j_{p}^\sigma-j_{ap}^\sigma)\over{\sum_\sigma j_{ap}^\sigma}}$$
where $j_{p(ap)}^\sigma$ is the current with spin $\sigma$ for
parallel (antiparallel) alignment of the magnetizations in the
magnetic layers. In Fig.~3(a) this TMR ratio is presented for an
asymmetric structure with different barrier heights as a function
of the thickness $b$. The TMR ratio reflects the oscillations of
the current density as well as the phase shift between the forward
and reverse bias (compare Fig.~2). It follows that the TMR ratio
is also very asymmetric  and the corresponding asymmetry ratio can
reach values up to $200$ (Fig.~3(b)) at the appropriate thickness
$b$.

This TMR asymmetry leads to a "magnetically controlled" diode,
whose blocking efficiency can be varied by a magnetic field. To
illustrate this, we calculate the value of relative
magnetoasymmetry (RMA)
\begin{equation}
\label{eq.4}
RMA={(j_{p}^{dir(inv)}/j_{p}^{inv(dir)})-(j_{ap}^{dir(inv)}/j_{ap}^{inv(dir)})\over{(j_{ap}^{dir(inv)}/j_{ap}^{inv(dir)})}}
\end{equation}
where $j_{p}^{dir(inv)}$ and $j_{ap}^{dir(inv)}$ are the total
forward (reverse) current for parallel and antiparallel
configuration of the magnetization, respectively. In Fig.~4 the
dependence of the RMA is shown as a function of $b$ for the double
barrier structure of~Fig.~3. At the maxima the magnitude of the
asymmetry ratio can be doubled by applying a magnetic field.

It is noted, that much higher asymmetries can be obtained when the
scattering in the metallic layers is weak. However, in this case
the resonance peaks become narrow and are therefore more difficult
to detect experimentally. More realistic is the case of strong
scattering which leads to a broadening of the resonance peaks and
consequently reduces the asymmetry ratios, with the advantage of
being easier to detect experimentally. More critical for possible
experimental observation of the predicted diode behaviour are
barrier and metallic spacer thickness fluctuations. Further
calculations show that thickness fluctuations which do not extend
over two atomic layers preserve the quantum well states. Details
of this work will be published elsewhere.
\begin{figure}
\begin{center}
\includegraphics[width=0.5\textwidth,angle=0]{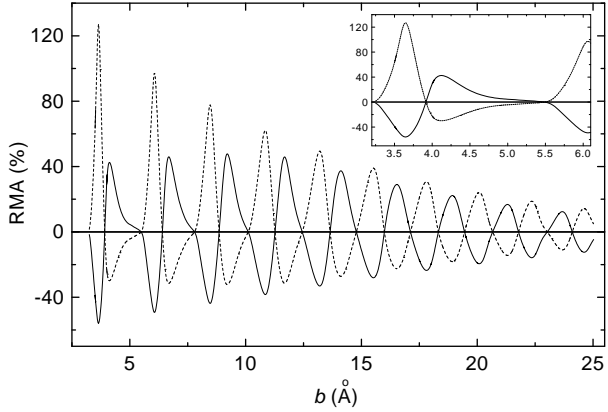}
\caption{The relative magnetoasymmetry (RMA) as a function of the
thickness $b$ for the same asymmetric double barrier structure as
for Fig.~3. The RMA is calculated from expression (\ref{eq.4})
with the current asymmetry ratio defined as
$(j_{p(ap)}^{dir}/j_{p(ap)}^{inv})$ (solid line) and
$(j_{p(ap)}^{inv}/j_{p(ap)}^{dir})$ (dashed line).}
\end{center}
\end{figure}
As a final point we would like to underline that the general
expressions~(\ref{eq.3}) describes properly $both$ ballistic and
diffuse tunneling regime through the system. The transition from
one regime to another can be retrieved if the scattering in the
middle metallic layer M$_3$ is zero ($l_3 \rightarrow \infty$). In
this case, the problem can be solved exactly and the analytic
expressions in~(\ref{eq.3}) yield that all tunneling probabilities
are equal ($D_{ij}^\sigma=D^\sigma$ for $i,j=2,4$) so that the
current density can be written as
$$j^\sigma=j_2^\sigma=j_4^\sigma=\displaystyle {e\over{\pi h}}\int
dE \big[f(E)-f(E+e\mathrm{V}_{ext})\big]\int
D^\sigma(E,\kappa)\kappa d\kappa $$ and the condition of constant
current across the whole double barrier structure is automatically
fulfilled. This result means that in the case of an ideal
structure without scattering, the purely quantum-mechanical
problem of electron tunneling through the $entire$ double barrier
structure is solved exactly describing the direct coherent
process. In this case, the voltage drop in each barrier is
proportional to its thickness as in refs.~\cite{Xiangdong}
and~\cite{Barnas}. This remarkable result is important since it
directly shows how the scattering in the middle layer destroys the
direct ballistic process so that the tunneling across the
structure is not described anymore by a single transmission matrix
but by resistors in series. Moreover, it means that the
expression~(\ref{eq.3}) can be written in the form:
\begin{equation}\label{eq.5}
\begin{array}{ll} j_{2(4)}^\sigma&=\displaystyle {e\over{\pi
h}}\int dE \big[f(E)-f(E+e\mathrm{V}_{ext})\big]\int
D_{24(42)}^\sigma(E,\kappa)\kappa d\kappa  \\
\\
&\displaystyle +{e\over{\pi h}}\int dE \big[f(E_1)-f(E_2)\big]\int
\left[D_{22(44)}^\sigma(E,\kappa)-D_{24(42)}^\sigma(E,\kappa)\right]\kappa
d\kappa
\end{array}
\end{equation}
where $E_1=E$, $E_2=E+e\mathrm{V}_{2}$ for $j_2$ and
$E_1=E+e\mathrm{V}_{2}$, $E_2=E+e\mathrm{V}_{ext}$ for $j_4$. The
first term in~(\ref{eq.5}) describes the direct ballistic process
across the structure and the second one describes local processes
which appear only in presence of the scattering in the middle
metallic layer.

In conclusion, we presented a quantum theory of the tunnel
magnetoresistance in magnetic double barrier structures and
predicted a strong "diode"-like behaviour for the \mbox{$I$-$V$}
characteristics and the TMR in the case of asymmetric barriers. It
was shown that the asymmetry ratio can be controlled by an applied
magnetic field. This phenomenon is due to the different phase
shift of the quantum well states in the middle metal layer under
forward and reverse applied voltage. This structure should have an
important application as a blocking device in a MRAM.

\acknowledgments The authors would like to thank Ursula Ebels for
a critical reading of the manuscript and Coriolan Tiusan for
fruitful discussions. This work was partially supported by the
European Brite Euram project 'Tunnelsense' (BRPR98-0657) and the
European IST project Nanomem (IST-1999-13471).


\begin{thebibliography}{0}

\bibitem{Moodera1}
J.S. Moodera, L.R. Kinder, T.M. Wong and R. Meservey, Phys. Rev.
Lett.{\bf 74}, 3273 (1995).

\bibitem{Moodera2}
J.S. Moodera and L.R.Kinder, J. Appl. Phys.{\bf 79}, 4724 (1996).

\bibitem{Miyazaki}
T. Miyazaki and N. Tezuka, J. Magn. Magn. Mater. {\bf 139}, L231
(1995).

\bibitem{Tezuka}
N. Tezuka and T. Miyazaki, J. Appl. Phys. {\bf 79}, 6262 (1996).

\bibitem{Parkin}
S.S.P.~Parkin {\it et. al.}, J. Appl. Phys. {\bf 85}, 5828 (1999).

\bibitem{Freitas}
S. Cardoso, P.P.~Freitas, C.~de~Jesus, P.~Wei  and J.C.~Soares,
Appl. Phys. Lett. {\bf 76}, 610 (2000).

\bibitem{Tehrani}
S.~Tehrani {\it et. al.}, J. Appl. Phys. {\bf 85}, 5822 (1999).

\bibitem{Xiangdong}
Xiandong Zhang, Bo-Zang Li, Gang Sun and Fu-Cho Pu, Phys. Rev. B
{\bf 56}, 5484 (1997).

\bibitem{Barnas}
M. Wilczynski, J. Barnas, J. Magn. Magn. Mater. {\bf 221}, 373
(2000).

\bibitem{Duke}
C.B. Duke, Tunneling In Solids, Academic Press, New York and
London, 1969.

\bibitem{Vedyayev2}
A. Vedyayev, N. Ryzhanova, C. Lacroix, L. Giacomoni and B. Dieny,
Europhys. Lett. {\bf 39}, 219 (1997).

\bibitem{Fisher}
D. Fisher and P. Lee, Phys. Rev. B {\bf 23}, 6851, (1981).

\bibitem{Camblong}
H. E. Camblong, P. M. Levy and S. Zhang, Phys. Rev. B {\bf 51},
16052 (1995).

\bibitem{Levy}
P. M. Levy ,H. E. Camblong   \and S. Zhang, J. Appl. Phys., {\bf
75}, 7076,6906 (1994).

\bibitem{Kane}
C. L. Kane, R. A. Serota and P. A. Lee, Phys. Rev. B {\bf 37},
6801 (1988).

\bibitem{Vedyayev4}
A. Vedyayev, N. Ryzhanova, R. Vlutters, B. Dieny and N. Strelkov,
J. Phys.: Condens. Matter {\bf 12}, 1797 (2000).

\bibitem{Kittel}
C. Kittel, Introduction to Solid State Physics, John Wiley $\&$
Sons, New York, 1986.

\bibitem{Bratkovsky}
A. M. Bratkovsky, Phys. Rev. B {\bf 56}, 2344 (1997).

\bibitem{Zhang}
S. Zhang and P. Levy, Phys. Rev. Lett. {\bf 81}, 5660 (1998).

\bibitem{Mathon}
J. Mathon \and A. Umerski, Phys. Rev. B {\bf 60}, 1117 (1999).

\bibitem{Vedyayev3}
A. Vedyayev, M. Chshiev, N. Ryzhanova \and B. Dieny, Phys. Rev. B
{\bf 61} 1366 (2000).

\bibitem{Moodera3}
J.S. Moodera {\it et. al.}, Phys. Rev. Lett. {\bf 83}, 3029
(1999).

\bibitem{Stearns}
M. B. Stearns, J. Magn. Magn. Mater. {\bf 5}, 167 (1977). The
choice of free electron-like parameters is justified by "ab
initio" calculations of the spin-dependent tunneling with
ferromagnetic 3$d$-metallic electrodes taking into account real
electronic structure (J. M. McLaren et al., Phys. Rev. B {\bf 56},
11827 (1997), W. H. Butler et al., Phys. Rev. B {\bf 63}, 054416
(2001)). It was shown that most of the tunneling current is
carried by hot spots in $\kappa$-space (in our case, the vicinity
of $\kappa=0$). This means that these electrons have a well
defined perpendicular component of the momentum which defines the
period of the current oscillations versus thickness $b$.


\end{thebibliography}
\end{document}